\documentclass[review]{elsarticle}

\usepackage{lineno,hyperref}
\usepackage{xcolor,amssymb}
\usepackage{amsmath}
\usepackage[normalem]{ulem}
\modulolinenumbers[5]

\journal{Journal of Molecular Liquids}

\begin{document}

\begin{frontmatter}

\title{Water under extreme confinement in graphene:  Oscillatory dynamics, structure, and hydration pressure explained as a function of the confinement width.}

%
%

\author[mymainaddress,mysecondaryaddress]{Carles Calero}
\ead{carles.calero@ub.edu}
\author[mymainaddress,mysecondaryaddress]{Giancarlo Franzese}
\ead{gfranzese@ub.edu}

\address[mymainaddress]{Secci\'o de
 F\'isica Estad\'istica i Interdisciplin\`aria--Departament de F\'{i}sica de la Mat\`{e}ria Condensada, Universitat de Barcelona, 08028 Barcelona, Spain}
\address[mysecondaryaddress]{Institut de Nanoci\`{e}ncia i Nanotecnologia, Universitat de Barcelona, Barcelona, Spain}

\begin{abstract}
Graphene nanochannels are relevant for their possible applications, as in water purification, and 
 for the challenge of understanding how they change the properties of confined liquids. Here, 
we use all-atom molecular dynamics simulations to investigate 
water confined in an 
open graphene slit-pore as a function of its width $w$, 
down to sub-nm scale.
We find that the water 
translational and rotational
dynamics exhibits 
an oscillatory
dependence on $w$, due to water   layering.
The oscillations in dynamics correlate with those in hydration pressure,
which  
can be negative (hydrophobic attraction),  or as high as 
 $\sim 1$ GPa, as seen in the experiments.  
At pore  widths commensurable with full layers
(around $7.0$ \AA\ and $9.5$ \AA\ for one and two layers, respectively),
the free energy of the system has minima, and 
the hydration pressure vanishes.  These are the separations at which the dynamics of confined water slows down.
Nevertheless, 
the hydration pressure vanishes also where the free energy has maxima, i.e., 
for those pore-widths which are incommensurable with the formation of
well-separated layers, as $w\simeq 8.0$ \AA. 
Around these values of $w$, the dynamics is
faster than in bulk, with water squeezed out from the pore. 
This behavior has not been observed for simple liquids under confinement, either for water in closed nano-pores. 
The decomposition of the free energy clarifies the origins of the dynamics speedups and slowdowns. In particular, we find that the nature of the slowdown depends on the number of water layers: for two  layers, it is due to the internal energy contribution, as in simple liquids, while for one layer, it has an entropic origin {\color{black} possibly} due to the {\color{black} existence of a hydrogen-bond network in water.} 
Our results shed light on the mechanisms ruling the dynamics and thermodynamics of confined water and are a guide for future  
experiments.
\end{abstract}


\end{frontmatter}


\section{Introduction}\label{Sect__{ in}tro}

Experiments and theories show that liquids in strong confinement  between solid boundaries exhibit very different dynamics and thermodynamics compared to the bulk \cite{Loche:2020aa, Knight:2019aa, Fumagalli:2018aa, Gao:2018aa, MCF2017, C7CP01962A, Nair2012, Falk:2010vn, Verdaguer:2006fk, Koga00}. The interaction of fluids with the confining surface  causes the structuring in  layers of the liquid  
 and affects its dynamics \cite{Slovak:1999bh, Bordin2012, Krott2013, Leoni_JCP2014, Leoni:2016aa, Bampoulis:2018aa, Tsimpanogiannis:2019aa}.
However, the experimental difficulties lead to debated results \cite{Radha:2016aa}.
For example,
a mobility of confined water higher than expected \cite{Majumder05, Holt2006, Majumder2011, Qin2011, Radha:2016aa} is confirmed by several  
  simulations of water  \cite{Mashl2003, Thomas:2008bh, Thomas2009, Qin2011, Ye2011, Barati-Farimani2011, Zheng2012},   and  of a water-like model \cite{Bordin2012}, 
 in  nanotubes  with a radius less than 1 nm. On the other hand, some simulations find that 
the diffusion coefficient of water in nanotubes simply decreases for decreasing confinement space down to 0.8 nm
\cite{Liu:2008vn, Nanok:2008uq}, pointing at  the relevance of the numerical limitations  \cite{Liu:2005hc, Mukherjee2007, Falk:2010vn, Kannam:2013aa}.

Hydrated graphene interfaces are of particular interest due to posible applications, including desalination \cite{Humplik:2011aa, Cohen-Tanugi:2012kx, OHern:2014kx,PhysToday-desal-2015, Kohler:2018ab} decontamination \cite{Romanchuk:2013fk, Vadahanambi:2013aa, Xu:2017aa}, energy storage \cite{Bo:2015aa, ISI:000453273200007}, heterogeneous catalysis, graphene exfoliation and transferring  \cite{Zangi2004, Bampoulis:2018aa}.
Graphene confinement enriches the complex phase diagram of bulk water 
in simulations~\cite{Zangi_PRL2003, Giovambattista_PRL2009, Gallo:2010bh, Mochizuki:2015ab, Corsetti_Artacho_PRL2016, PhysRevLett.116.025501, Roman:2016aa, Ruiz-Pestana:2018aa, Raju:2018aa, Gao:2018aa, Abbaspour:2018aa, Engstler:2018aa}
 and experiments \cite{Algara-Siller:2015aa, Calo_Verdaguer_JPhysChemC2015, Daio2015, Rouziere:2018aa},
and changes the water dynamics and structure, 
as seen numerically \cite{Kumar07, cicero2008, PhysRevB.78.075432, Gordillo:2010aa, Falk:2010vn, Boukhvalov:2013fk, Zhao:2015ab, Mozaffari:2016aa, MCF2017, C7CP01962A, Jiao:2017aa, Chakraborty:2017aa, Neek-Amal:2018aa}
and in laboratories \cite{Nair2012, Joshi:2014fk, Zhou:2018aa}.

Experiments  in slit pores show that water forms layers parallel to the walls \cite{Iiyama:1995aa}. Computer simulations of water between two nanoscopic hydrophobic plates reveal that these layers correspond to local minima in the free-energy profile as a function of the plate-plate separation $w$ \cite{Choudhury:2005aa}
and that 
the water {\color{black}mobility}  in a hydrophobic slit pore monotonically increases
as $w$  becomes larger \cite{Choudhury:2005ij, Marti_JPCB2006,Giovambattista:2006vn, Nagy_JPCB2007, Mosaddeghi_JCP2012, Sanghi_JCP2013, Chialvo_JPCC2013, Bordin_JPCB2013, Diallo_PRE2015, Mendoncca_JCP2020}. Similar results hold  for water confined in graphite \cite{Hirunsit:2007aa} and quartz \cite{Zangi_PRL2003, Giovambattista_PRL2009, Giovambattista:2006vn}, with 
freezing of the dynamics at sub-nm confinement.

As the  slit pore size $w$ decreases,  experiments and simulations  show
the emergence of oscillatory behavior in solvation forces and  viscosity of water
\cite{Verdaguer:2006fk,  Neek_ACSNano2016, Engstler:2018aa, Engstler:2018ab}.
This behavior is reminiscent of the oscillations in the diffusion constant and the free energy  of 
 a simple (Lennard-Jones) liquid confined in a slit-pore \cite{Gao_PRL1997}, and 
it calls into question a recent numerical work finding a plate-plate attraction that monotonically increases
for decreasing $w$ \cite{Samanta:2018aa}. 
It is, therefore, unclear  how strong (sub-nm) pore confinement affects water diffusion, hydration forces, and free energy, and which relation holds  among them.

To contribute to this debate,
here, we use all-atom molecular dynamics (MD) simulations to investigate the dynamical, structural, and thermodynamic properties of water confined between rigid graphene plates  as a function of their separation,
 $w$, down to sub-nm size.  
To properly account for the most relevant conditions in which confined water is encountered in nature and in possible applications, we couple the confined system to a large reservoir of water molecules which are kept at constant temperature and pressure. We demonstrate that, under such conditions, the translational and rotational water dynamics have 
 oscillatory
dependence on the plate separation, and, through hydration-pressure and free-energy calculations,  we show that such behavior directly correlates with structural properties of the confined water. 
 The free-energy analysis allows us to clarify the origins of the dynamics speedups and slowdowns. In particular, we find that the  slowdown mechanism depends on the number of water layers: it is due to the entropy when there is only one layer, while it is caused by the internal energy when the layers are two.

\begin{figure}[]
\begin{center}
\vspace*{0.5cm}
\includegraphics[angle=0, width=6.0cm]{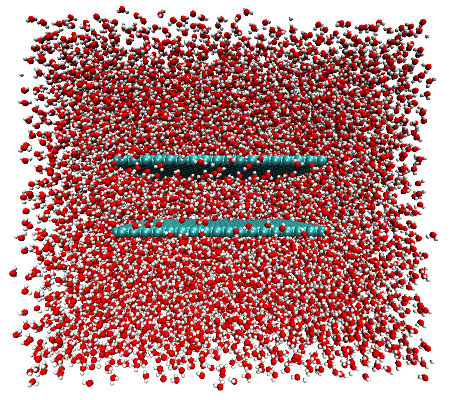}
\caption{Our simulation box with two parallel graphene-plates surrounded by water. 
Carbon atoms are black, oxygen atoms red, and hydrogen atoms white. 
For clarity purposes, we do not show all the water molecules. 
 The pore width $w$ is the distance between the plates.
}
\label{Fig:snapshot}
\end{center}
\end{figure}

\section{Methods}\label{Sect__Simulations}

We simulate a system made of two $24.6$ \AA $ \times 25.5$ \AA\ rigid graphene-plates surrounded by water with periodic boundary conditions in the three spatial directions (Fig.~\ref{Fig:snapshot}). We consider fixed  slit-pore widths $6 \leq w/\text{\AA} \leq 17$
and focus on the water molecules  between the plates. 

We use TIP4P/2005-water~\cite{Abascal_Vega_JCP2005}
and carbon atoms interacting through a Lennard-Jones potential 
 with parameters from the  CHARMM27 force field. 
We deduce the parameters of the water-carbon interaction 
 adopting the Lorentz-Berthelot rules, and cut off the Van
der Waals interactions at 12 \AA\ with a smooth switching function
starting at 10 \AA. We compute the long-range electrostatic forces using
the particle mesh Ewald method \cite{Essmann_JCP1995} with a grid space
of $\approx 1$ \AA.  We use the GROMACS package~\cite{Gromacs4} at  constant temperature $T$ and volume $V$, with 1 fs 
simulation time-step, and update the electrostatic interactions every 2 fs.  If not indicated otherwise, we fix $T=300$ K and, 
following other authors, e.g.,
\cite{Radha:2016aa, Giovambattista_PRL2009, Giovambattista:2006vn, Giovambattista09}, we
adopt a Berendsen thermostat~\cite{Berendsen_JPhysChem1984} to control it. 

Thermostating highly confined fluids has no perfect solution
\cite{Bernardi:2010aa,Kannam:2011aa, Kumar-Kannam:2012aa}, and the Berendsen thermostat does not accurately reproduce the canonical fluctuations of temperature, which are relevant for thermodynamic properties, such as the specific heat.  However, comparison of velocity rescaling, Berendsen, or Nos\'e-Hoover thermostats shows no effects on layering, diffusion, and the thermodynamic averages of confined water, which are the objects of our study \cite{Loche:2020aa, Zaragoza_PCCP2019}. 
We fix the total volume of our simulation box to $V=4.2\times 4.2\times 5.1$ nm$^3$ and the total number of water molecules to $N=2796$.

We prepare the initial configurations, corresponding to the different slit-pore width, with the help of the visualization software VMD \cite{VMD}. First, we create the graphene layers and keep the positions of the carbon atoms frozen in the three spatial directions during the simulations. Then, we hydrate the system with TIP4P/2005 water molecules. 
We perform a short 1000-steps minimization run  to avoid overlaps between molecules. Next, we simulate the system at constant $N$, $V$, and $T$, equilibrating for 5 ns. Then, we collect data every 10 ps for the next 50 ns and every 0.1 ps for the next  8 ns \cite{Gromacs4}.

To avoid plates-boundaries effects, we analyze water molecules within the $A \times w$ central region between the plates, where $A = 15 \text{ \AA} \times 15 \text{ \AA}$. 
Because this confined sub-region is open and connected to a reservoir of water at constant $N$, $V$, and $T$, as it is the case of many experimental setups \cite{Nair2012}, the observables are calculated at constant chemical potential, $\mu$, constant pressure $P$, and constant $T$. As we will discuss, allowing the density of the confined region to fluctuate leads to important differences with studies of similar confined systems simulated at a fixed density \cite{cicero2008, Zhao:2015ab}.

\section{Results and Discussion}

\subsection{Translational Dynamics}

First, we study the dynamics --both translational and rotational-- of water molecules confined between graphene plates at fixed $w$. 
To avoid biased sampling of the dynamics, we consider all water molecules within the central volume $A \times w$ between the plates and collect their trajectories while they remain in the region. To improve statistics in the calculation of the evolution of the mean square displacement, we use multiple time origins.

\begin{figure}[h!]
\begin{center}
\vspace*{0.cm}
\includegraphics*[angle=0, width=7.5cm]{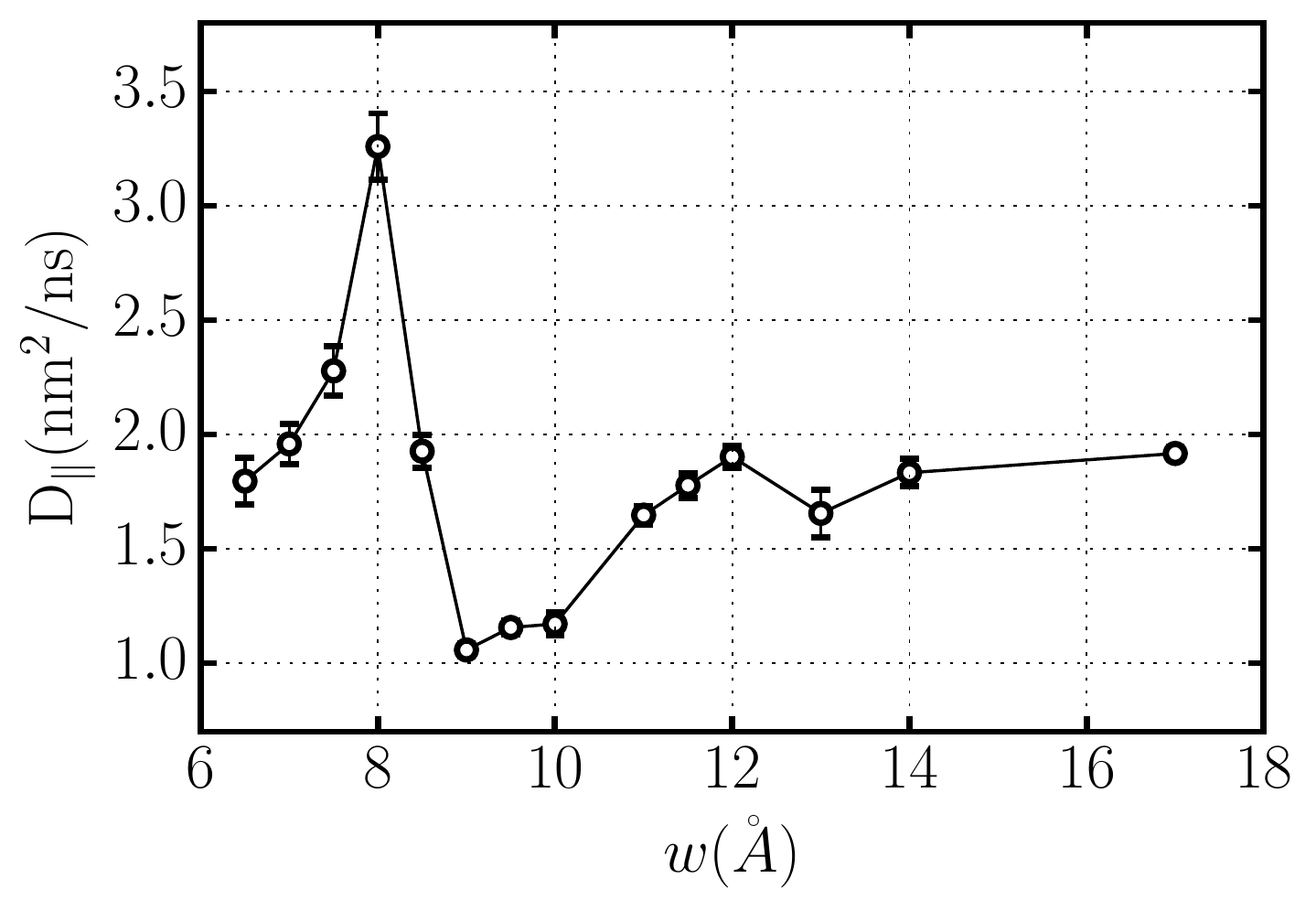}
\caption{The confined-water diffusion coefficient $D_{\parallel}$, parallel to the plates,  shows 
oscillatory
behavior as a function of the plate separation $w$,
with maxima at $w\simeq 8$~\AA\ and 12~\AA, and minima at $w\simeq 9$~\AA\ and 13~\AA
 }
\label{Fig:diff300}
\end{center}
\end{figure}

We find that $D_{\parallel}$, the diffusion coefficient of confined water along the directions parallel to the plates,  exhibits 
oscillatory
 behavior as a function of  $w$ (Fig.~\ref{Fig:diff300}).
In particular, $D_{\parallel}$ has  two maxima and, at least, two minima for $6.5 \leq  w/\text{\AA}\leq 17$, with the largest maximum, $\simeq 3.3$~nm$^2$/ns, at  $w\simeq 8$~\AA\ and the smallest value, $\simeq 1.0$~nm$^2$/ns, at $w\simeq 9$~\AA. 
For large $w$, and outside the confined region,  the diffusion coefficient approaches the bulk value  for TIP4P/2005-water, $\approx 2.1$~nm$^2$/ns at ambient conditions \cite{Abascal_Vega_JCP2005}.

Hence, the translational diffusion of water in a slit-pore slows down,  
by a factor $\approx 0.5$
for $w = 9\text{ \AA}$, with respect to bulk water, while speeds up, by a factor 
$\approx 1.65$, when the confining distance is $w = 8\text{ \AA}$, and finally decreases again for smaller values of $w$.
The speedup  is reminiscent of the debated \cite{Thomas:2008bh, Falk:2010vn, Kannam:2013aa} enhanced mobility of water in 
pores of a few nanometers \cite{Holt2006, Majumder05, Majumder2011, Qin2011, Radha:2016aa}{\color{black}, or at the water-vapor interface \cite{Fabian:2016aa}}.

Also, we analyze the  {\em survival probability}, i.e., the probability that a water molecule spends a time $t$ in the confined region, defined as
\begin{equation}
 S^{(w)}(t) \equiv \left\langle \frac{N^{(w)}(t_0, t_0 + t)}{N^{(w)}(t_0)} \right\rangle\, ,
\end{equation}
where $N^{(w)}(t_0, t_0 + t)$ is the number of water molecules which remain in the confined region after a time $t$ out of those that were in that region at $t_0$, $N^{(w)}(t_0)$. %
The brackets $\langle ... \rangle$ indicate average over different time origins $t_0$. 

\begin{figure}[h!]
\begin{center}
\vspace*{0.cm}
\includegraphics*[angle=0, width=7.5cm]{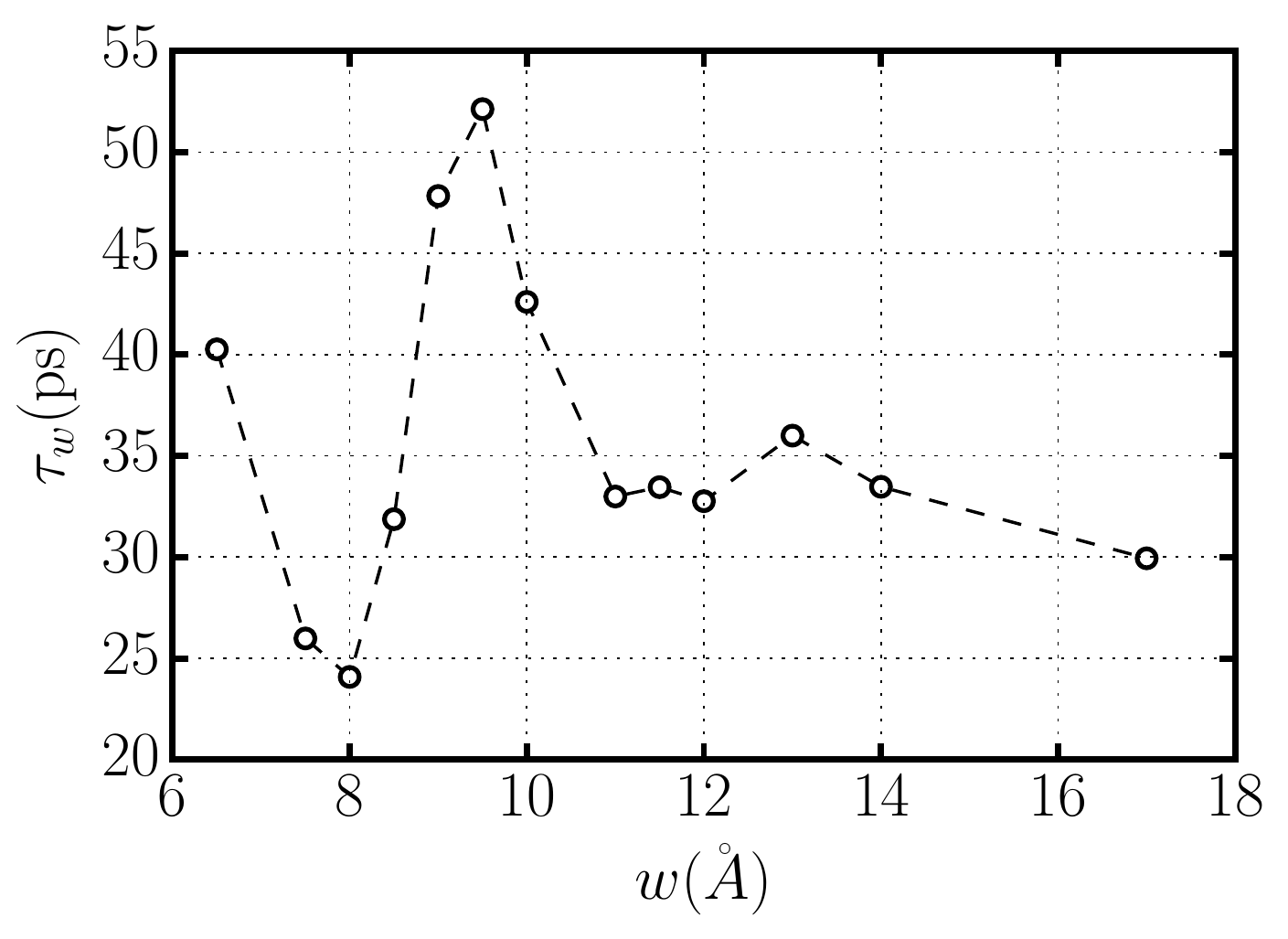}
\caption{ Characteristic time of water occupancy of the confined region $\tau_w$ as a function of the plate separation $w$.}
\label{Fig:Tsurv}
\end{center}
\end{figure}

{\color{black} We find that $S^{(w)}$ decays slowly and can be adjusted with a double exponential function, or a single exponential for $t>1$ ps.}
The  time  $\tau_w$ at which $S^{(w)}(\tau_w) = 0.5$ is another property that characterizes the confined water, and its behavior (Fig.~\ref{Fig:Tsurv}) confirms the oscillatory dependence of the translational dynamics  as a function of the plate separation $w$, reaching minima where $D_{\parallel}$ has maxima and {\it vice versa}\footnote{ The average thermal diffusion velocity along the main axis of the pore is of the order of $10^{-3}$ m/s, that is, as expected, at least  two orders of magnitude smaller than the  flow velocity of water  under large pressure gradients (1 PPa/m) through graphene nanocapillaries \cite{Radha:2016aa}.}.

\subsection{Rotational Dynamics}

\begin{figure}[h!]
\begin{center}
\vspace*{0.cm}
\includegraphics*[angle=0, width=7.5cm]{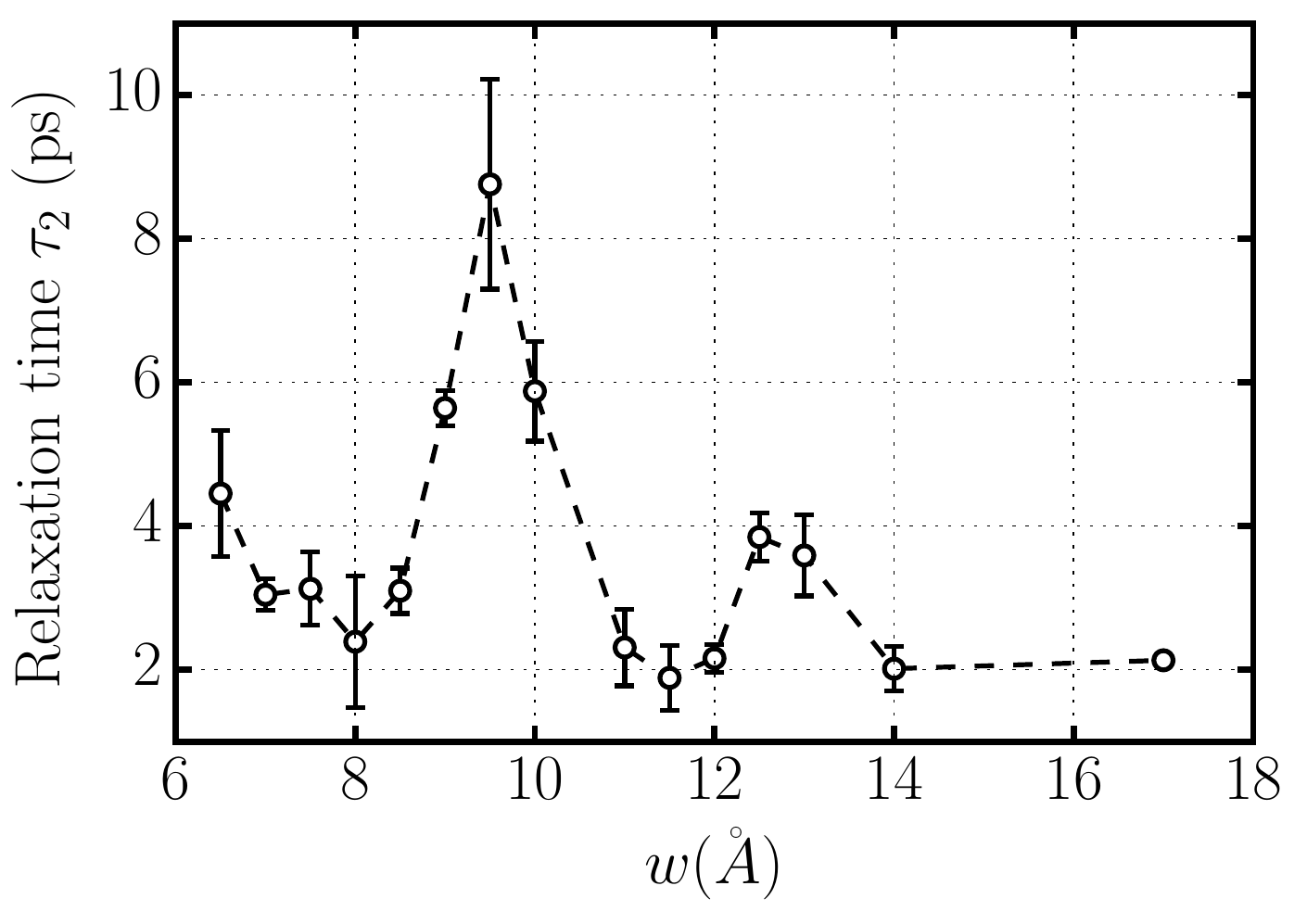}
\caption{The reorientation correlation time $\tau_2$ of confined water as a function of graphene-plate separation $w$ is oscillatory with minima and maxima that correlates well with those of translational $\tau_w$.}
\label{Fig:RotCorrFunc}
\end{center}
\end{figure}

Next, we probe the effect of confinement on the rotational dynamics of water molecules by calculating the rotational dipolar correlation functions,  
\begin{equation}
 C_{n}(t) \equiv \langle P_n\left(\hat{\mu}(t)\cdot \hat{\mu}(0)\right) \rangle\,,
\end{equation}
where $P_n(x)$ is the $n$-th order Legendre polynomial,  $\hat{\mu}(t)$ is the direction of the water dipole vector at
time $t$ and $\langle ... \rangle$ denote ensemble average over all
water molecules and time origins. To quantify the
relaxation of the correlation functions $C_{n}(t)$, we define
the relaxation time 
\begin{equation}
 \tau_n \equiv \int_0^{\infty} C_{n}(t) dt\,,
\label{tau}
\end{equation}
which is not dependent on any assumptions on the functional form of the correlation function. 

We  find (Fig.~\ref{Fig:RotCorrFunc}) that {\color{black} for $w\geq 14$ \AA} the rotational dynamics of water molecules, characterized by the reorientation correlation time $\tau_2$, is $\tau_2\approx (2.0 \pm 0.2)$ps, consistent with values for bulk TIP4P/2005-water at $T =300$ K \cite{Calero_JPhysChemB2015}. For $w<14$ \AA, also $\tau_2$ exhibits an oscillatory dependence on the distance $w$. It has, at least, two maxima, reaching a value $(\approx 8.5\pm 1.5)$ps, i.e.,  
$\approx 4.25$ times larger than bulk, 
at $w \simeq 9.5\text{ \AA}$. It decreases to the bulk value for $w \simeq 8\text{ \AA}$, and finally increases again for smaller separations. 
The maxima and minima of $\tau_2$ correlate well with those of the translational $\tau_w$ and the oscillatory behavior of  $D_{\parallel}$, suggesting a coupling of translation and orientation dynamics in confined water under the thermodynamic conditions considered here, and consistent with the existence of cooperative rearranging regions due to the water hydrogen bonds \cite{delosSantos2012}.

\subsection{Structure}

\begin{figure}[h!]
\begin{center}
\vspace*{0.cm}
\includegraphics*[angle=0, width=7cm]{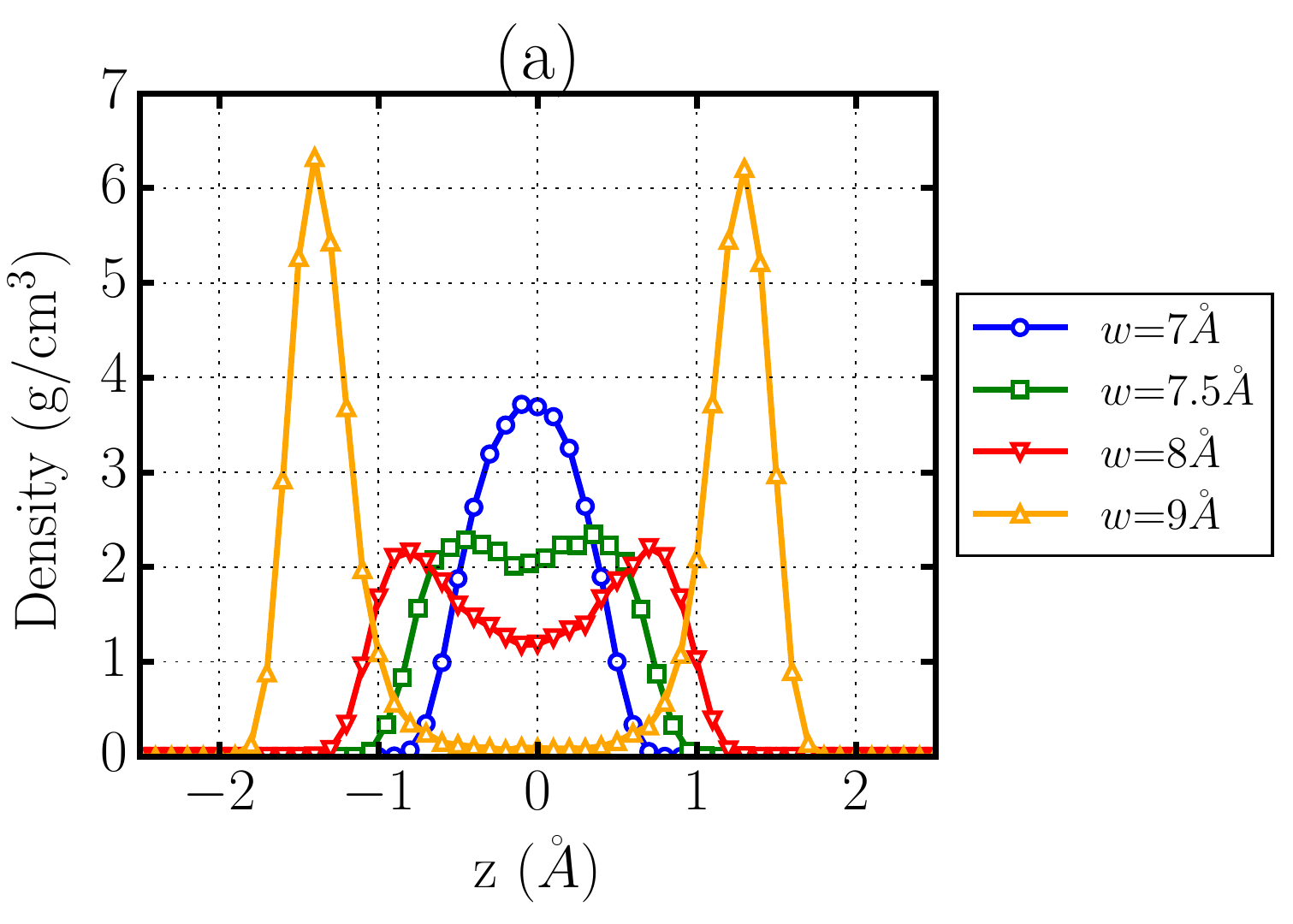}
\includegraphics*[angle=0, width=7cm]{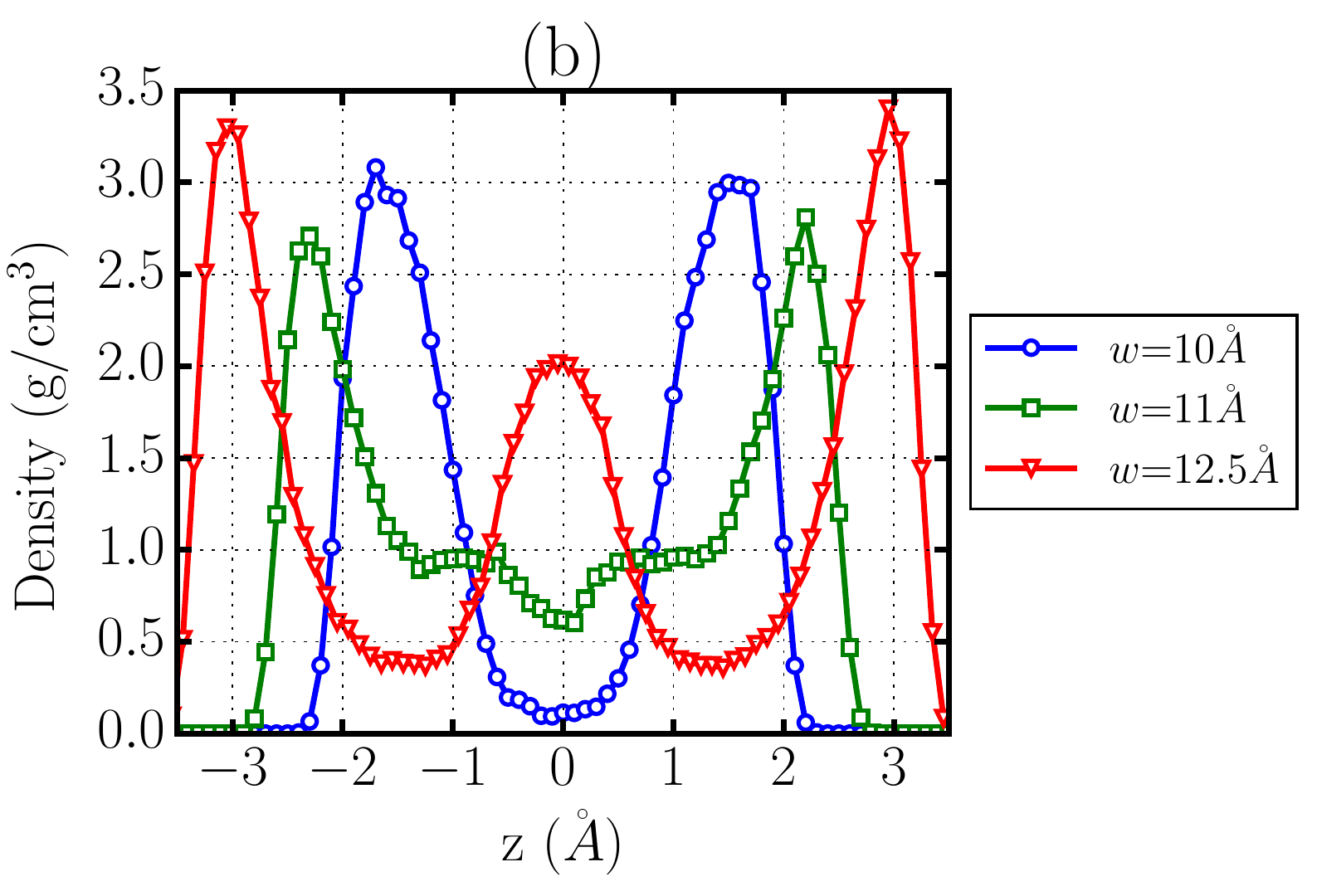}
\includegraphics*[angle=0, width=7cm]{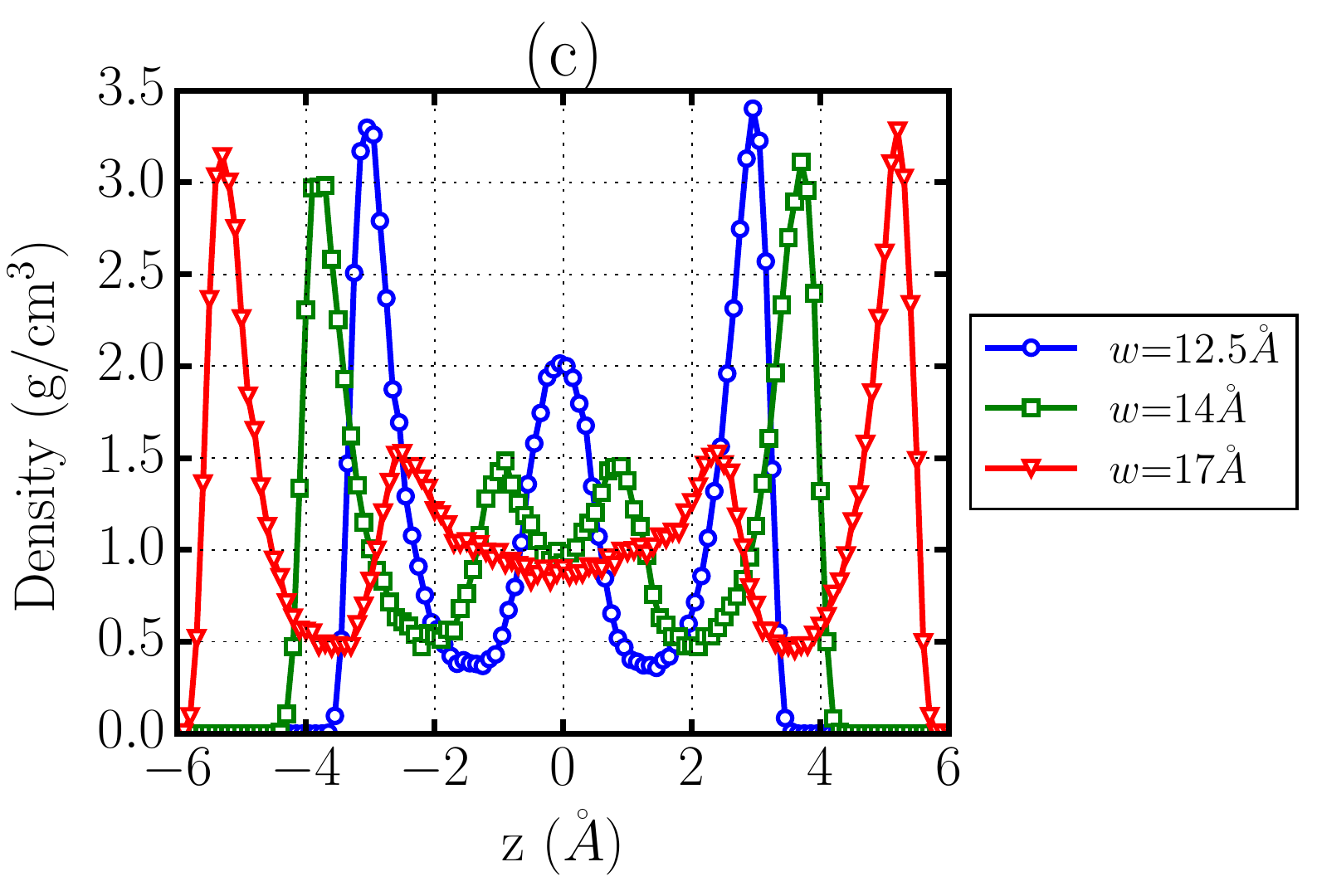}
\caption{Water density-profile as a function of the distance $z$ from the center of the slit-pore, for different values of graphene-plate distance $w$.  We find: (a)  from one to two layers for 
$7.0 < w/\text{\AA}\leq 9.0$,
(b)  from two to three layers for 
$10.0 < w/\text{\AA}\leq 12.5$, 
(c) from three to four layers for 
$12.5 < w/\text{\AA}\leq 17.0$. }
\label{Fig:watdens}
\end{center}
\end{figure}

For all the plate separations we considered here, we find that the confined water remains in the liquid phase and 
 organizes into layers parallel to the confining walls at well defined values of $w$: one layer for $6.5 \leq w/\text{\AA}\leq 7.0$, two  for {\color{black} $9.0 < w/\text{\AA}< 10.0$, three  for $12.5 < w/\text{\AA}< 14.0$} (Fig.~\ref{Fig:watdens}). 
 These results are consistent with previous works performed under similar conditions \cite{Engstler:2018ab}\footnote{
 At pore width 9 \AA, we find a peak height for the density profile that is approximate twice the values in Ref.s \cite{cicero2008} and \cite{Zhao:2015ab}.  This apparent discrepancy  occurs because  these Ref.s adopt the constant-$NVT$ ensemble, while we use the constant-$\mu PT$ ensemble, as in Ref. \cite{Engstler:2018ab}.  Also, this observation is supported by the results in Ref. \cite{Gao_JPhysChemB1997}, where the authors find the same peak height for Lennard-Jones particles with size  3.405 \AA, comparable to a water molecule, in a slit-pore with 9 \AA\ width, simulated at constant-$\mu PT$ in a geometry similar to the one we adopt here.
  }
\footnote{{\color{black}
Layering occurs at any interface, including  at liquid-vapor interfaces once the smearing effect of the capillary waves is removed \cite{Fabian:2016aa}
}}   .

\begin{figure}[h!]
\begin{center}
\vspace*{0.cm}
(a) \includegraphics*[angle=0, width=7.5cm]{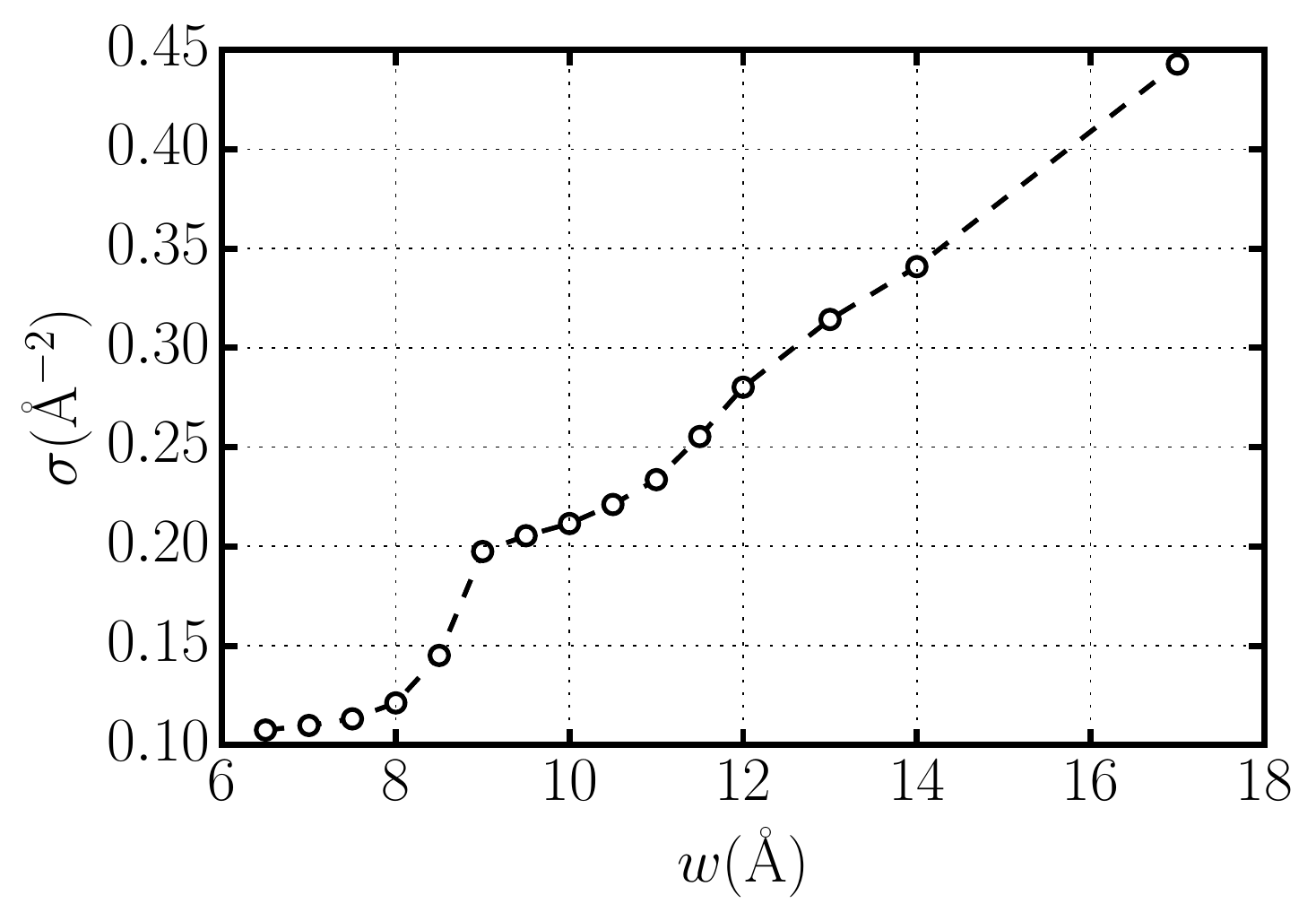}\\
(b) \includegraphics*[angle=0, width=7.5cm]{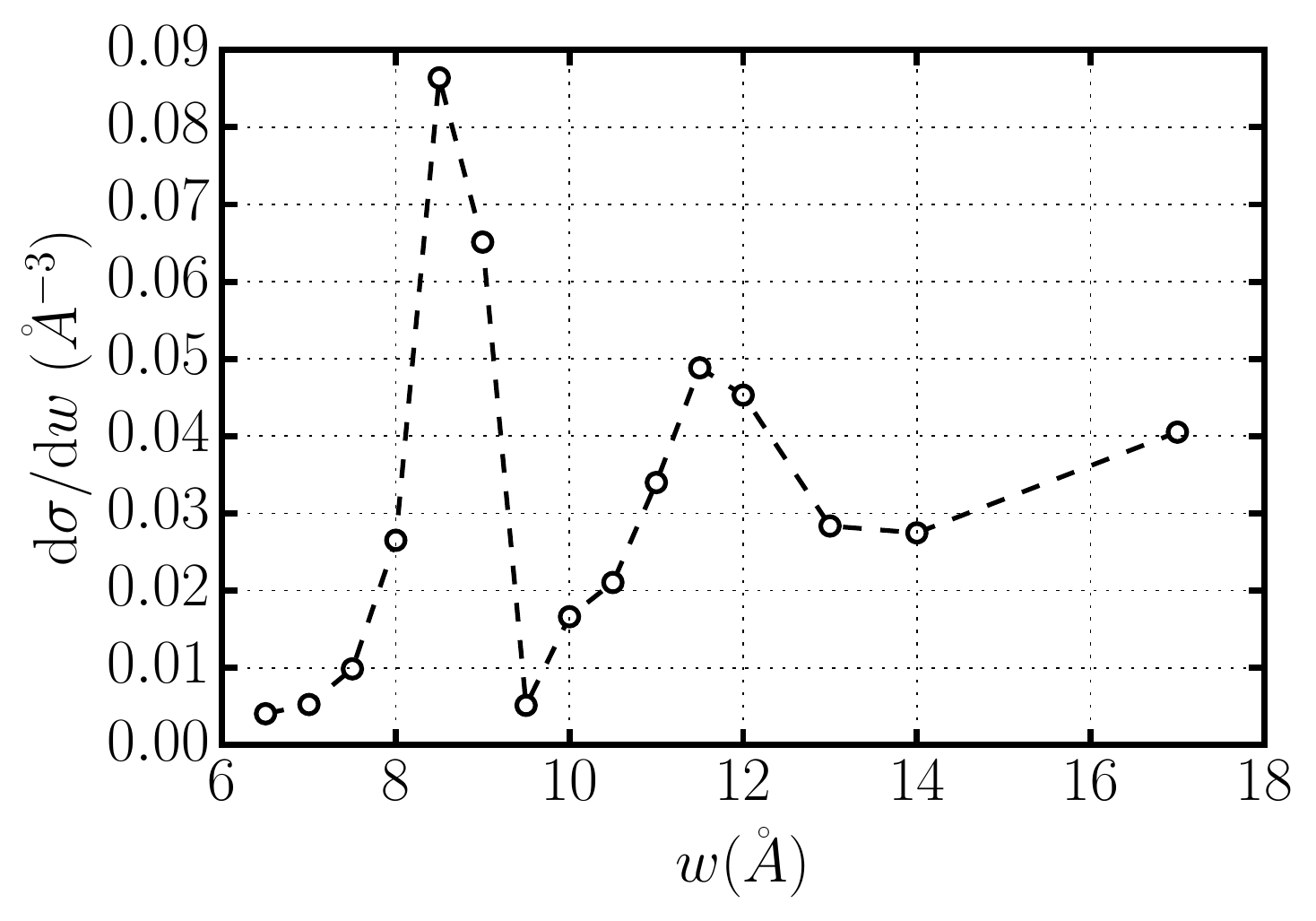}
\caption{The {\color{black}  slit-pore acceptance capacity per unit area}
$\sigma$ and its derivative as a function of graphene-plate separation $w$.
(a) $\sigma$ increases monotonically with $w$, but at different rates depending on $w$.
(b) The derivative $d\sigma/dw$ is non-monotonic, with maxima and minima. The derivative reaches values close to zero at $w\approx 7.0\text{ \AA}$ and $9.5\text{ \AA}$.}
\label{Fig:Nwat}
\end{center}
\end{figure}

As a result of the formation of layers, the 
{\color{black}  slit-pore {\it acceptance capacity} per unit area} $\sigma\equiv N(w)/A$ 
does not depend linearly with plate separation $w$  but exhibits a steplike pattern (Fig.~\ref{Fig:Nwat}a),
consistent with previous results \cite{Engstler:2018ab}, 
 as emphasized by plotting the derivative  
$d\sigma/dw$ (Fig.~\ref{Fig:Nwat}b). Here,  $N(w)$ is the number of water molecules confined in the central region $A \times w$ and is calculated by integrating the density profiles at each $w$.
The derivative
 $d\sigma/dw$ is a measure of the {\color{black} variation of the acceptance} capacity of the slit-pore 
upon changing $w$ and exhibits a non-monotonic behavior, as expected due to the layering. 
Minima in 
$d\sigma/dw$
occur at values of $w$ at which full layers are present. 
Maxima in $d\sigma/dw$, instead, occur 
for those values of $w$ such that a new layer is forming.

We observe that 
 the non-monotonic behavior of $d\sigma/dw$ shows a clear correlation with the dependence of confined water dynamics with plate separation $w$. Indeed, both translational (Fig.~\ref{Fig:diff300}) and rotational (Fig.~\ref{Fig:RotCorrFunc}) water dynamics are faster when 
 {\color{black} $d\sigma/dw$ is  larger}, and become slower when $d\sigma/dw$ decreases. Our observation corroborates the conclusion from experiments showing a  flow enhancement  in channels with only a few layers of water, that the authors associate with an increased structural order in nanoconfined water \cite{Radha:2016aa}.

\subsection{Hydration pressure}

To understand the relation of the dynamics with the thermodynamics, we calculate the {\it hydration pressure}, i.e., the pressure exerted by confined water perpendicularly to the graphene plates, defined as 
\begin{equation}
\Pi \equiv p_{\textrm{in}}^{\perp} - p_ {\textrm{out}}^{\perp},
\label{pi}
\end{equation} 
where $p_ {\textrm{in}}^{\perp}$ and $p_ {\textrm{out}}^{\perp}$ are the internal and, respectively, external pressures of water in the direction perpendicular to the plates, calculated as the vectorial sum over all the forces due to water acting on the plates, divided by the section (surface) of the plates  \cite{Varnik_JCP2000, Giovambattista_PRL2009}. 

\begin{figure}[h!]
\begin{center}
\vspace*{0.cm}
\includegraphics*[angle=0, width=7.5cm]{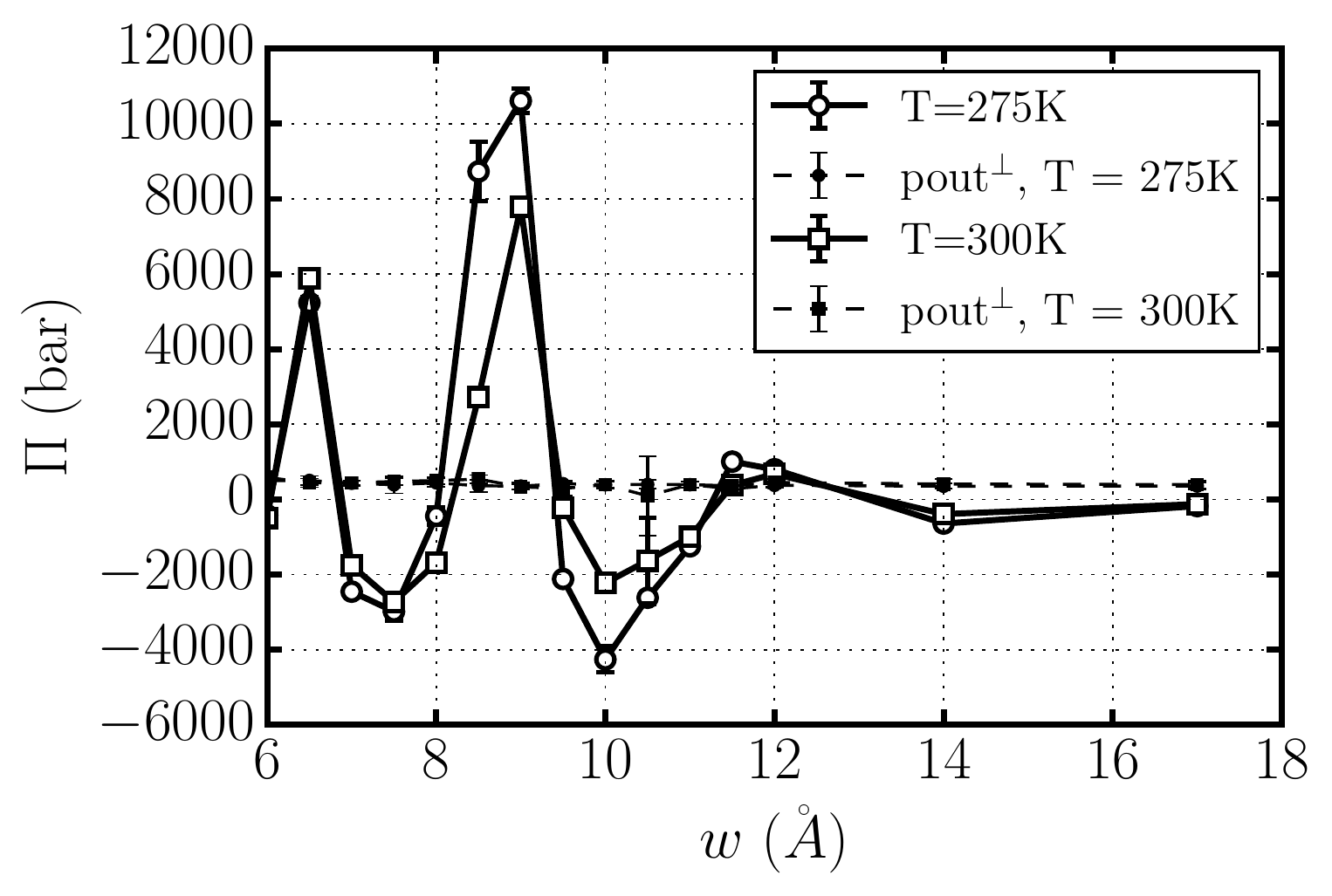}
\caption{Average pressures as a function of graphene-plate separation $w$ at $T=300$ K (squares) and 275 K (circles).
In both cases, the hydration pressures $\Pi$ (continuous lines)  are non-monotonic, while the external pressures (dashed lines) are constant. 
The external pressure $p_ {\textrm{out}}^{\perp} = (400 \pm 100)$ bar is constant with respect to $w$.
Where not visible, the error bars are smaller than the symbols' size. 
}
\label{Fig:hydforce}
\end{center}
\end{figure}

We calculate $\Pi$ for each  specific values of $w$ and for two different temperatures, 300 K and 275 K (Fig.~\ref{Fig:hydforce}).
Due to the relatively small size of the confined region with respect to the simulated system, the external pressure $p_ {\textrm{out}}^{\perp}$ is insensitive to the change in the slit-pore width $w$ and is given by $p_ {\textrm{out}}^{\perp} = (400 \pm 100)$ bar.

We find that $\Pi$ oscillates between positive and negative values. 
The values of $w$ at which the  $\Pi$ minima and maxima occur do not display evident dependence on $T$ \cite{Engstler:2018ab} and are at distances separated by approximately the molecular diameter of the water molecule. 
Minima and maxima follow a pattern that resembles that of dynamical quantities (Figs. \ref{Fig:diff300}, \ref{Fig:Tsurv}, \ref{Fig:RotCorrFunc}) and the 
{\color{black} acceptance capacity}
variation (Fig. \ref{Fig:Nwat}) with $w$, but do not coincide. 

In particular, we observe that it is $\Pi \approx 0$ at values that are close to those  at which $D_{\parallel}$, $\tau_w$,  $\tau_2$ and $d\sigma/dw$ have local maxima or minima. 
Comparison with the water density-profile (Fig. \ref{Fig:watdens})
clarifies that {\color{black} some of} the values of $w$ at which $\Pi \approx 0$ correspond to those for optimal layer-wall, and  layer-layer, separations. 
For example, at $T=300$ K this value is $6.5<w/\text{\AA}< 7.0$ for one layer, for two layers is 
{\color{black} $9.0< w/\text{\AA}< 10.0$. However,  $\Pi$ vanishes also at intermediate values of $w$, which are, as we will discuss next, related to marginally-stable configurations}.

At intermediate values of $w$, the amount of confined water could be above or below the number that optimizes the inter-layers distance, causing repulsion ($\Pi>0$) or attraction ($\Pi<0$), respectively, between the plates. 
In these cases, an external constraint on the plates' position provides the mechanical stability along the direction perpendicular to the slit-pore.

At $T=300$ K, the necessary external pressure to keep the plates at a fixed distance is $\approx 0.6$ GPa for $w=6.5\text{ \AA}$, a width smaller than the optimal value for one confined water monolayer, and $\approx 0.8$ GPa for at $w=9\text{ \AA}$, which is  just below the optimal value for two layers.
Instead, 
the walls effectively attract each other 
(with $\approx -0.3$ GPa for $w=7.5\text{ \AA}$, 
and $\approx -0.2$ GPa for $10.0\text{ \AA}$)
for $w$ above the optimal 
distance for one or two water layers.

Lowering $T$ at 275 K,
the effective plate-plate attraction  increases to $\approx -0.4$ GPa for $w=10.0\text{ \AA}$, and 
the repulsive pressure increases to $\approx 1.1$ GPa for $w=9\text{ \AA}$.
This large variation of $\Pi$ at $T=275$ K
is due to the energy change associated to the reentrant crystallization into a bilayer hexagonal ice for $9.0<w/\text{\AA}<9.5$ \cite{MCF2017}, that is larger than the energy change for 
 restructuring liquid water at $T=300$ K.
These results are consistent, for order of magnitude, with the 
capillary pressure estimated in simulations  of  graphene pores
 that accommodate two 
 layers of water  ($\approx  1$ GPa at  $\simeq 9$  \AA)
\cite{Algara-Siller:2015aa, He:2019aa}.
 
Hence, our calculations confirm that (i) water-mediated plate-plate forces oscillate \cite{Engstler:2018ab, Engstler:2018aa}, at variance with Ref. \cite{Samanta:2018aa}; (ii) the 
oscillations correlate with large structural changes of the solvent, e.g., the layers merging or the ice melting.
Furthermore, we clarify that the water dynamics oscillates as well, with speedups or slowdowns, with respect to bulk water, associated to these large pressures changes.
To better understand the thermodynamic nature of these dynamic changes, we analyze next the free energy of the confined water.

\subsection{Free energy}

\begin{figure}[h!]
\begin{center}
\vspace*{0.cm}
\includegraphics*[angle=0, width=7.5cm]{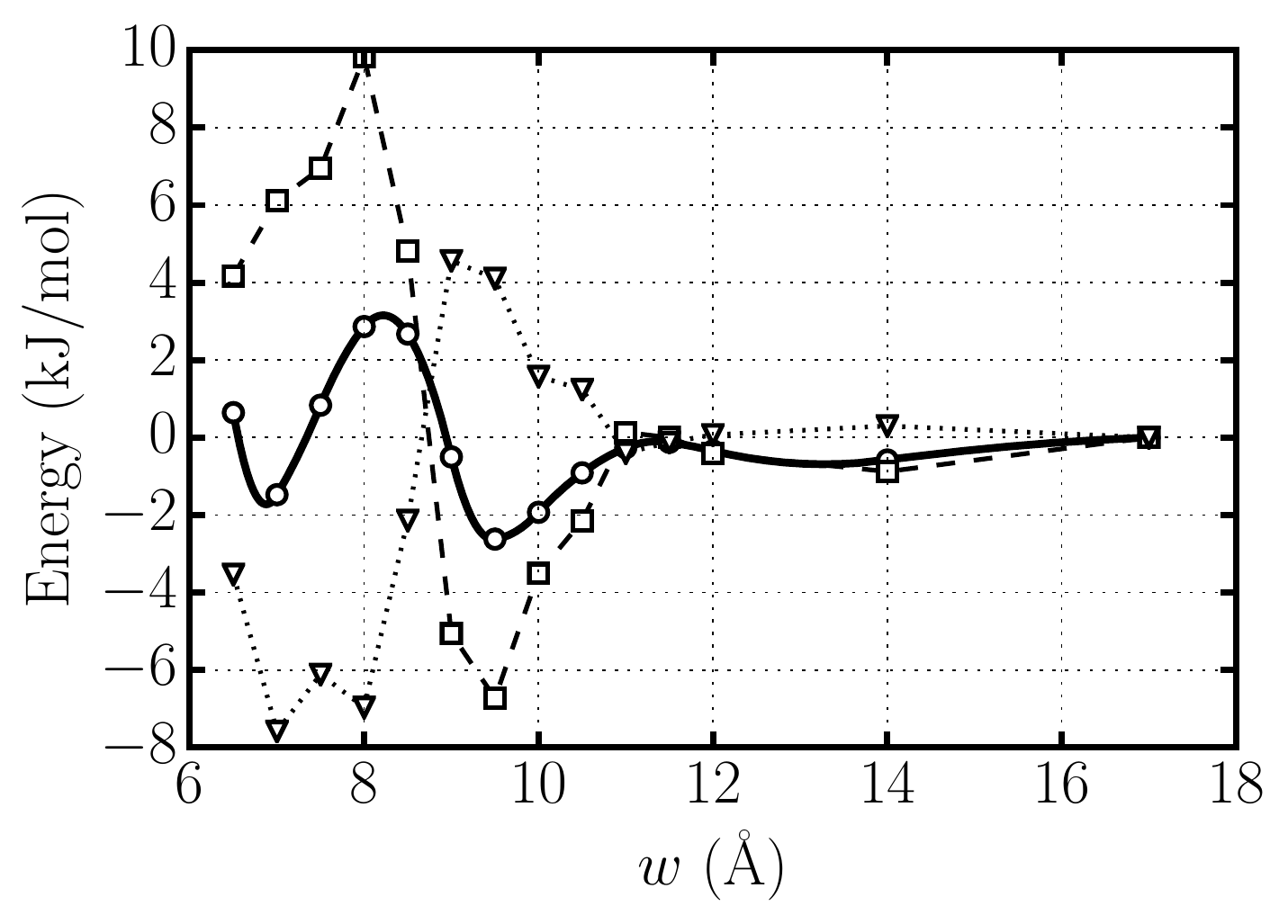}
\caption{Work per confined water molecule performed to approach the graphene plates from $w_0=17\text{ \AA}$ to $w$, $\Delta f(w)$ (solid line with circles),  as a function of the
separation $w$ at $T=300$ K. 
This work has two contributions: the internal energy part (dashed line with squares) and the entropic part (dotted line with with triangles). The minimum of $\Delta f(w)$ at $w \approx 9.5\text{ \AA}$ (water bilayer) corresponds to a minimum in internal (attractive) energy $\Delta u(w)$ per water molecule, while the minimum at $w \approx 7.0\text{ \AA}$ (water monolayer) to a minimum in the entropy term (disordering) $-T\Delta s_\textrm{in}$ per water molecule, despite the positive internal (repulsive) energy.
}
\label{Fig:free_energy}
\end{center}
\end{figure}

Following Gao et al. ~\cite{Gao_PRL1997, Gao_JPhysChemB1997}, we  calculate the  
 free-energy variation per confined water molecule, $\Delta f(w)$, as 
the work done against the hydration forces to approach the two graphene plates
 from   our largest separation,  $w_0=17\text{ \AA}$, to any smaller $w$,  
 where 
  \begin{equation}\label{Eq:F}
 \Delta f(w)\equiv  \Delta f(w_0\rightarrow w) \equiv f(w)-f(w_0)  \equiv  -\int_{w_0}^{w} \mathfrak{f}_{ \textrm{hyd}}^{\perp}(w')dw'\,,
\end{equation}
 and
$\mathfrak{f}_{\textrm{hyd}}^{\perp}(w)\equiv \Pi(w) /\sigma(w)$ is the hydration force, per confined water molecule, acting on the plates at separation $w$ (Fig.~\ref{Fig:free_energy}).
 
Our calculation of $\Delta f(w)$ clarifies that the work per molecule necessary to approach the two confining surfaces is non-monotonic with $w$, as well as all the other dynamic, structural and thermodynamic quantities we presented here.
In particular, the work oscillates depending on how strong is the water-mediated effective interaction between the  walls, with extrema in $\Delta f(w)$ correlated to 
{\color{black} the zeros of $\Pi$, with minima and maxima in $\Delta f(w)$ corresponding to stable and marginally-stable configurations,  respectively}. 

The negative values of $\Delta f(w)$  show that the confined system gains energy at the  $w$ of the minima. 
In particular, the work per molecule has its absolute minimum,
$\Delta f(w)\approx -3.0$ kJ/mol,
 at $w \approx 9.5\text{ \AA}$, i.e., when there are two layers of confined water (Fig.~\ref{Fig:watdens}a). 
Another strong minimum, 
$\Delta f(w)\approx -1.8$ kJ/mol,
occurs at $w \approx 7.0\text{ \AA}$, corresponding to one single layer of confined water, while the minimum,
$\Delta f(w)\approx -0.6$ kJ/mol,
 at $w \approx 13.0\text{ \AA}$, with three confined layers, is weaker.
These results, showing that nanoconfined water forms preferentially monolayers and bilayers,
 are consistent with experiments  
under similar conditions ~\cite{Xu_Science2010, Calo_Verdaguer_JPhysChemC2015, Santos_Verdaguer_Materials2016} and recent simulations\cite{Engstler:2018ab}.

To understand the nature of  minima we calculate the separate component of the free energy. 
First, we observe that Eq. (\ref{Eq:F}) corresponds to the Helmholtz  free energy of confined water in a volume with a constant number of water molecules \cite{Gao_PRL1997, Gao_JPhysChemB1997}.
Hence, 
$\Delta f\equiv \Delta f_\textrm{in}= \Delta u_\textrm{in} -T\Delta s_\textrm{in}$, 
where 
$\Delta u_\textrm{in}\equiv u_\textrm{in}(w)-u_\textrm{in}(w_0)$  
is the internal energy change per confined water molecule, and 
$\Delta s_\textrm{in}\equiv s_\textrm{in}(w) - s_\textrm{in}(w_0)$ 
is the corresponding entropy change per confined water molecule.
Because in our system $N(w)\ll N$, and the water outside the pore is, in first approximation, not affected by the change in $w$, we approximate  
$\Delta u_\textrm{in}\approx \Delta u$ 
of the entire system, simplifying the calculation for the long-range electrostatic interaction  (requiring periodic boundary conditions for the  Ewald summation).

Knowing $\Delta f$ and $\Delta u$, we estimate $-T\Delta s_\textrm{in}$ as their difference (Fig.~\ref{Fig:free_energy}).
We find that the nature of the two free-energy minima for water bilayer and monolayer are quite different.

The free-energy minimum at  $w \approx 9.5\text{ \AA}$ (water bilayer) is dominated by the  internal  energy contribution $\Delta u<0$ among the water molecules, inducing an effective attractive force between the walls for $9.5 \leq w/$\AA$\leq 11.5$, and an effective repulsion for $8.0 < w/$\AA$< 9.5$ (Fig.\ref{Fig:hydforce}).
The entropy term $-T\Delta s_\textrm{in}>0$ is associated with an increase of order in the confined water bilayer with respect to the multilayers at $w_0$.
This is similar to what has been found in confined Lennard-Jones liquids \cite{Gao_PRL1997}.

The free-energy minimum at $w \approx 7.0\text{ \AA}$ (water monolayer)  is dominated, instead, by the 
 entropy contribution $-T\Delta s_\textrm{in}$, i.e., the monolayer has a structure that is more disordered than the water multilayer. Furthermore, the internal energy  $\Delta u>0$ among the water molecules is overall repulsive. Despite this repulsion, the effective  force between the walls is attractive for $7.0 \leq w/$\AA$\leq 8.0$ and repulsive for $6.0 < w/$\AA$< 7.0$ (Fig.\ref{Fig:hydforce}).  
{\color{black} This effect could be a consequence of the hydrogen bond interactions among water molecules.}
The strong confinement, indeed, induces a deformation of the hydrogen bonds  \cite{Zangi_PRL2003, Algara-Siller:2015aa}, increasing {\color{black} both} the energy of the  {\color{black}  configurations} ($\Delta u>0$) and  {\color{black} their} degeneracy ($\Delta s_\textrm{in}>0$).
   
The two free-energy minima are separated by a maximum, $\Delta f(w)\approx 3.0$ kJ/mol, at $w\approx 8\text{ \AA}$. This {\color{black} free-energy barrier} is due to the combination of the stronger water-water repulsion  ($\Delta u>0$)  and the  disordering effect ($\Delta s_\textrm{in}>0$) associated with the expulsion of water from the pore (Fig.\ref{Fig:watdens}){\color{black}, and its  value},  $\approx 5.0$ kJ/mol, 
is twice as large as the thermal energy per mole at $T=300$ K. 

\section{Summary and Conclusions}

We study the dynamics and thermodynamics of water confined between two graphene walls at distance $w$, 
 down to the sub-nm scale. Our molecular dynamics simulations show that all the calculated quantities  have oscillatory dependence  on $w$ due to layering.

We focus on the translational and rotational dynamics. We find that (i) the diffusion constant $D_\parallel$  parallel to the confining walls, (ii) the characteristic time $\tau_w$ of occupancy of the pore, and (iii) the reorientation correlation time $\tau_2$ of confined water oscillate, as a function of the plate separation $w$, and correlate among them. 

The  overall dynamics slows down at pore widths commensurable with three, one, and, in particular, two 
full layers. It speeds up, instead, at pore widths that are incommensurable with full layers, with a surprisingly large factor at sizes between a bilayer and a monolayer. Squeezing the pore around 8 \AA, considerably increases the thermal diffusion  of the confined water, supporting the enhanced mobility seen in experiments with carbon nanotubes, membranes, and capillaries with nm-sized pores   \cite{Holt2006, Majumder05, Majumder2011, Qin2011, Radha:2016aa}.

To the best of our knowledge, this oscillatory sequence of repeated minima and maxima for translational and rotational dynamics of water, in a graphene slit pore  as a function of the plate separation, has not been reported so far. Consistent numerical results  by 
Neek-Amal et al. \cite{Neek_ACSNano2016} show oscillatory shear viscosity of water  between two parallel graphene layers, separated by less than 2 nm.  As in our analysis, 
the oscillations originate from the commensurability between the capillary size and the size of water molecules. 
If the Stokes-Einstein (SE) relation were valid in extreme confinement, the $D_\parallel$ oscillation  
would be related to those of the viscosity.
However,  K\"ohler et al. \cite{Kohler_PCCP2017} show that the SE relation breaks in narrow (with diameters less than 4.07 nm) hydrophobic nanotubes, questioning the validity of the SE also in hydrophobic slit-pores, and, as a consequence, the possible relation between shear viscosity and diffusion coefficient in nm-size graphene confinement.

On the other hand, previous results for Lennard-Jones particles in a slit-pore, with particle sizes comparable to water molecules, 
display a diffusion constant that oscillates below confining distances of 2 nm \cite{Gao_PRL1997}. As in our case, Gao et al. relate this behavior to oscillating hydration forces and oscillating free energy. However, differently from our case, they find no dynamics speedup in confinement with respect to the bulk.  The fast diffusion and rotation dynamics in very narrow pores 
{\color{black} could be}, therefore, a peculiar property of water, that we can explain in terms of hydration forces and free energy.

In particular, our free energy analysis elucidates the origin of the water behavior in sub-nm confinement.
When the water  bilayer forms at $w \approx 9.5\text{ \AA}$, the water-water overall attraction, and the structural ordering, generate a free-energy minimum that slows down the dynamics. At the pore width for a full bilayer, the hydration pressure is zero, and there is no need for external forces to reach mechanical stability. 

By reducing $w$, water repulsion and structural disorder take over. At  $w \approx 8.0\text{ \AA}$, it is necessary to apply high external pressure (of the order of $\approx 1$ GPa) to reach mechanical stability for the confined water, and the water dynamics largely speeds up. 

Below $w \approx 8.0\text{ \AA}$, the wall-wall effective interaction is again attractive and the system collapses toward the width  $w \approx 7.0\text{ \AA}$, corresponding to the free-energy minimum for a  confined monolayer. 
At the same time, the dynamics slows down to the bulk value approximately.
The origin of the monolayer free-energy minimum is, however,  quite different from the bilayer case. 
It is the increase of entropy of the monolayer, with respect to bulk,  that generates the minimum, while 
the overall water-water interaction is repulsive at this confinement. {\color{black} We suggest that this result could be } an effect of $a$) the distortion of the water hydrogen bonds and $b$) the consequent large degeneracy within 
 water configurations with the same internal energy. 

Consistent with experimental observations, we find that the bilayer at $w=9.5$ \AA\  is more stable than the monolayer at $\approx 7.0$ \AA~\cite{Calo_Verdaguer_JPhysChemC2015, Xu_Science2010, Santos_Verdaguer_Materials2016} and that the transformation from bilayer to monolayer requires an energy that is twice as large as the thermal energy per mole at $T = 300$~K. The same transformation  is associated to a reentrant crystallization into a bilayer hexagonal ice at $T = 275$~K.

We find that, within this sub-nm range of pore widths, the hydration-pressure 
oscillations are much stronger than at larger pore sizes.
This result clarifies that the apparent contradiction between recent numerical works, with  
\cite{Engstler:2018aa, Engstler:2018ab} or without oscillations \cite{Samanta:2018aa} in the water-mediated  wall-wall interaction, is possibly due to a lack of resolution in the sub-nm range of the pore size.
Furthermore, we expect increases in the compressibility of the confined water at those pore widths at which the hydration pressure is zero but the system is mechanically 
{\color{black} marginally-stable}, i.e., near the free-energy maxima at 
$8\leq w/\text{\AA}\lesssim 8.5$ and $w \approx 11.5\text{ \AA}$, consistent with Ref. \cite{Engstler:2018ab}.

In conclusion, our 
results provide a thermodynamically consistent account of recent experimental observations  for water confined in  graphene slit pores, or similar confinements.
They shed light on the origin of oscillations of the hydration force under sub-nanometer confinement from a structural and a thermodynamical perspective, resolving apparent contradictions in recent results. Furthermore, our results reveal, for the first time, the oscillatory behavior of the dynamical properties of confined water on the same scale. We 
{\color{black} suggest} that the nature of such oscillations is a unique
feature of sub-nm confined  water, supporting further studies for possible applications.

\section{Acknowledgments}
We thank Fabio Leoni and Jordi Mart\'i 
for useful discussions.  We acknowledge the support of Spanish grant PGC2018-099277-B-C22 
(MCIU/AEI/ERDF). 
CC acknowledges the support from the
Catalan Governament Beatriu de Pin\'os program (BP-DGR 2011). 
GF acknowledges the support by ICREA Foundation (ICREA Academia prize).
This work was partially funded by Horizon 2020 program through 766972-FET-OPEN-NANOPHLOW. 
The authors thankfully acknowledges the computer resources, technical
expertise and assistance provided by the Red Espa\~nola de
Supercomputaci\'on.

\section*{References}

\bibliography{Bibliography}

\end{document}